\documentclass{aastex631}

\begin{document}

\title{Evidence for Plasmoid-mediated Magnetic Reconnection during a Small-scale Flare in the Partially Ionized Low Solar Atmosphere}

\author{Guanchong Cheng}
\affiliation{Yunnan Observatories, Chinese Academy of Sciences,Kunming, Yunnan 650216, P. R. China}
\affiliation{University of Chinese Academy of Sciences, Beijing 100049, P. R. China}
\author{Lei Ni}
\affiliation{Yunnan Observatories, Chinese Academy of Sciences,Kunming, Yunnan 650216, P. R. China}
\affiliation{University of Chinese Academy of Sciences, Beijing 100049, P. R. China}
\affiliation{Center for Astronomical Mega-Science, Chinese Academy of Sciences, \\
20A Datun Road, Chaoyang District, Beijing 100012, P. R. China}

\author{Zehao Tang}
\affiliation{Yunnan Observatories, Chinese Academy of Sciences,Kunming, Yunnan 650216, P. R. China}
\affiliation{University of Chinese Academy of Sciences, Beijing 100049, P. R. China}

\author{Yajie Chen}
\affiliation{Max Planck Institute for Solar System Research, Justus-von-Liebig-Weg 3, 37077, Göttingen, Germany}

\author{Yuhao Chen}
\affiliation{Yunnan Observatories, Chinese Academy of Sciences,Kunming, Yunnan 650216, P. R. China}
\affiliation{University of Chinese Academy of Sciences, Beijing 100049, P. R. China}

\author{Jialiang Hu}
\affiliation{Yunnan Observatories, Chinese Academy of Sciences,Kunming, Yunnan 650216, P. R. China}
\affiliation{University of Chinese Academy of Sciences, Beijing 100049, P. R. China}

\author{Jun Lin}
\affiliation{Yunnan Observatories, Chinese Academy of Sciences,Kunming, Yunnan 650216, P. R. China}
\affiliation{University of Chinese Academy of Sciences, Beijing 100049, P. R. China}
\affiliation{Center for Astronomical Mega-Science, Chinese Academy of Sciences, \\
20A Datun Road, Chaoyang District, Beijing 100012, P. R. China}







\begin{abstract}

Magnetic reconnection plays a crucial role in the energy release process for different kinds of solar eruptions and activities. The rapid solar eruption requires a fast reconnection model. Plasmoid instability in the reconnecting current sheets is one of the most acceptable fast reconnection mechanisms for explaining the explosive events in the magnetohydrodynamics (MHD) scale, which is also a potential bridge between the macroscopic MHD reconnection process and microscale dissipations. Plenty of high resolution observations indicate that the plasmoid-like structures exist in the high temperature solar corona, but such evidences are very rare in the lower solar atmosphere with partially ionized plasmas. Utilizing joint observations from the Goode Solar Telescope (GST) and the Solar Dynamics Observatory (SDO), we discovered a small-scale eruptive phenomenon in NOAA AR 13085, characterized by clear reconnection cusp structures, supported by Nonlinear Force-Free Field (NLFFF) extrapolation results. The plasmoid-like structures with a size about 150 km were observed to be ejected downward from the current sheet at a maximum velocity of 24 km$\cdot$s$^{-1}$ in the H$\alpha$ line wing images, followed by enhanced emissions at around the post flare loop region in multiple wave lengths. Our 2.5D high-resolution MHD simulations further reproduced such a phenomenon and revealed reconnection fine structures. These results provide comprehensive evidences for the plasmoid mediated reconnection in partially ionized plasmas, and suggest an unified reconnection model for solar flares with different length scales from the lower chromosphere to corona. 

\end{abstract}



\section{Introduction} \label{sec:intro}

Magnetic reconnection is a key process in plasma physics that converts magnetic energy to thermal energy, bulk kinetic energy and accelerates particles, which frequently happens in the laboratory experiments, the Earth's magnetosphere, the solar atmosphere, and other astrophysical environments \citep{Kopp1976, Priest1982}. Various activities at different length scales in the universe are related with magnetic reconnection, such as solar flares and Coronal Mass Ejections (CMEs) in the sun's atmosphere, magnetic storms in the Earth's magnetosphere, and high-energy events in cosmic space \citep[e.g.,][]{Hirayama1974, Parker1988, Rubenstein2000, Moore2001, Qiu2004, Angelopoulos2008, Miyake2013}. The basic physic mechanisms in magnetic reconnection have been intensively studied in the past few decades\citep{Priest2000, Ji2022}.\\

As we know, the classical steady state Sweet-Parker\citep{Parker1957,Sweet1958} and Petschek\citep{Petschek1964} models face significant challenges for explaining the dramatic explosive energy release process in the solar atmosphere. Theories, along with numerical experiments based on MHD, Hall MHD and Particle-in-Cell (PIC) models indicate that the current sheet with high Lundquist number can usually be fragmented due to plasmoid instability across different length scales \citep[e.g.,][]{Biskamp1993, Biskamp2000, Priest1986, Loureiro2007,Bhattacharjee2009,Daughton2009,Shepherd2010}. Without invoking kinetic mechanisms, the plasmoid instability in MHD scale can still lead to a high reconnection rate $\sim$ 0.01 \citep{Bhattacharjee2009, Ni2015}, which weakly depends on the Lundquist number. Therefore, the plasmoid instability is considered as one of the main fast reconnection mechanisms for explaining the large scale explosive events in the solar and astrophysical environments. The theoretical analysis \citep{Shibata2001} and high resolution numerical studies \citep{Ni2015} demonstrate that the plasmoid instability can cascade from the large MHD scale down to the kinetic scale, serving as a key bridge among physical processes across different scales. The macroscopic current sheet and microscopic energy dissipation processes are connected through this mechanism \citep{Ji2022}.  \\

Numerous observational studies have provided evidences suggesting the presence of plasmoid instability in the solar corona. \cite{Ohyama1997} detected the plasma ejection before the impulsive phase of solar flares with two-ribbons by using the soft and hard X-ray observations, which indicate the occurrence of rapid magnetic reconnection and plasmoid instability. \cite{Lin2005} utilizing instruments such as Ultraviolet Coronagraph Spectrometer (UVCS) and Large Angle and Spectrometric Coronagraph (LASCO) on the Solar and Heliospheric Observatory (SOHO) spacecraft, they detected the clear plasmoid-like blobs in the long current sheets behind CMEs, and the estimated reconnection rate is between 0.01 and 0.23. The bidirectional moving plasmoid-like blobs along the current sheet, the interactions and merging of these blobs, and their collisions with the post flare loop were also detailedly captured by the SDO \citep[e.g.,][]{Takasao2012}. The bright blobs in the reconnection current sheet between a filament and a corona loop \citep{Li2016}, along a corona jet \citep[e.g.,][]{ZhangQM2014,ZhangQM2019} have also been detected. Recent ultra-high-resolution observations have also detected blobs, approximately on the scale of 1 Mm with temperatures around 2 MK, moving bidirectionally along the fan-spine magnetic field structure during the coronal reconnection events \citep{ChengX2023}.\\

Both single-fluid and multi-fluids MHD simulations show that the plasmoid instability can also occur in the partially ionized plasmas \citep[e.g.,][]{Leake2012,Ni2015,Cheng2021}, and result in the fast reconnection in the low solar atmosphere \citep{Ni2015,Ni2018}. The turbulent magnetic reconnection driven by plasmoid instability can result in the mixing of cool and much hotter plasmas, which may account for the simultaneous occurrences of multi-wavelength brightenings in a single event \citep{Ni2021, ZhangQM2016, ZhangQM2019}. The recent advanced radiative MHD simulations and synthesized spectral line profiles within different wave bands prove that the hot ultraviolet (UV) bursts and the much cooler Ellerman bombs (EBs) can coexist in a turbulent reconection region, even within the same plasmoid in the lower solar chromosphere \citep{Ni2021, Cheng2024}.\\

However, direct observational evidences of plasmoid instability in the partially ionized low solar atmosphere remain relatively scarce, primarily due to the much smaller size of the reconnection event in this region and the limited resolutions of the existing solar telescopes. \cite{Yang2015} and \cite{Xue2016}, observed small-scale current sheets with a length of a few Mm using the high temporal-spatial resolution H$\alpha$ data from the New Vacuum Solar Telescope, the non-uniform structures were detected, even though distinct plasmoid structures were not evident. EBs \citep{Ellerman1917} and UV bursts \citep{Peter2014}, are the observed smallest reconnection events in the solar atmosphere so far, they usually exhibit as the flame-like structures in the H$\alpha$ wing image and the small bright point structures in the Si IV image, respectively. It is very difficult to discern their corresponding current sheets and fine structures inside \citep{Tian2016,Young2018,Chen2019}. More recently, \cite{Rouppe2023} utilized unique spectral observations from the Microlensed Hyperspectral Imager (MiHI) on the Swedish Solar Telescope to study the EBs and they found the ejected plasmoid-like structures in the H$\alpha$ wing image. Nevertheless, confirming the existence of plasmoid instability in the partially ionized lower solar atmosphere, investigating their relations with the small activities and finding out the common characteristics with large scale activities in the solar corona, still require further high resolution observations and more in-depth studies.\\

We conducted researches on a  reconnection activity in active region 13085 using GST's H$\alpha$ images combined with SDO/AIA images, along with our high-resolution 2.5D MHD simulation. In this work, we confirmed the presence of plasmoid instability in a current sheet observed in H$\alpha$ wing image and the interactions between plasmoids with the small post flare loop in the region of interest (ROI), providing comprehensive evidences for the existence of plasmoid instability in the lower solar atmosphere. The observational and simulation results are presented in Section 2. We finish with a summary and  discussion in Section 3.\\

\begin{figure}[ht!]
\plotone{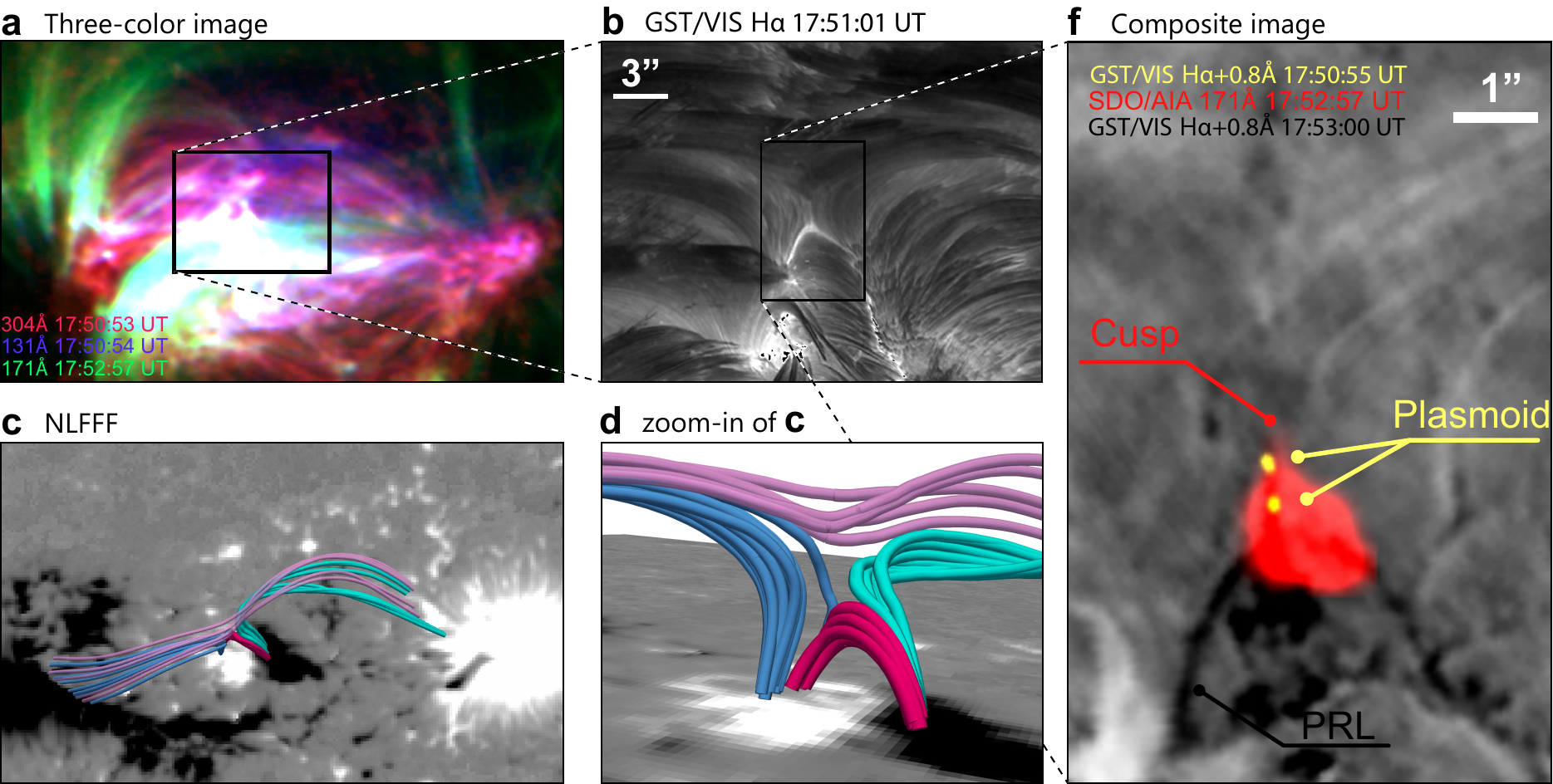}
\caption{Overview of the ROI. (a): tri-color image; the red, purple and green components are AIA 304 \AA, 131 \AA\ and 171 \AA\ images, respectively. (b): the VIS H$\alpha$ observation of the reconnection region; the field of view (FOV) is indicated by the box in (a). (c): the NLFFF extrapolation of the ROI; the FOV is the same with (a). (d): the zoom-in of (c). (f): the composite image; the yellow and gray components are two VIS H$\alpha$+0.8 \AA\ images at two different time, while the red is the AIA 171 \AA\ image. The FOV of (f) is indicated by the box in (b).
\label{fig:general}}
\end{figure}

\begin{figure}[ht!]
\begin{interactive}{animation}{ani.mp4}
\plotone{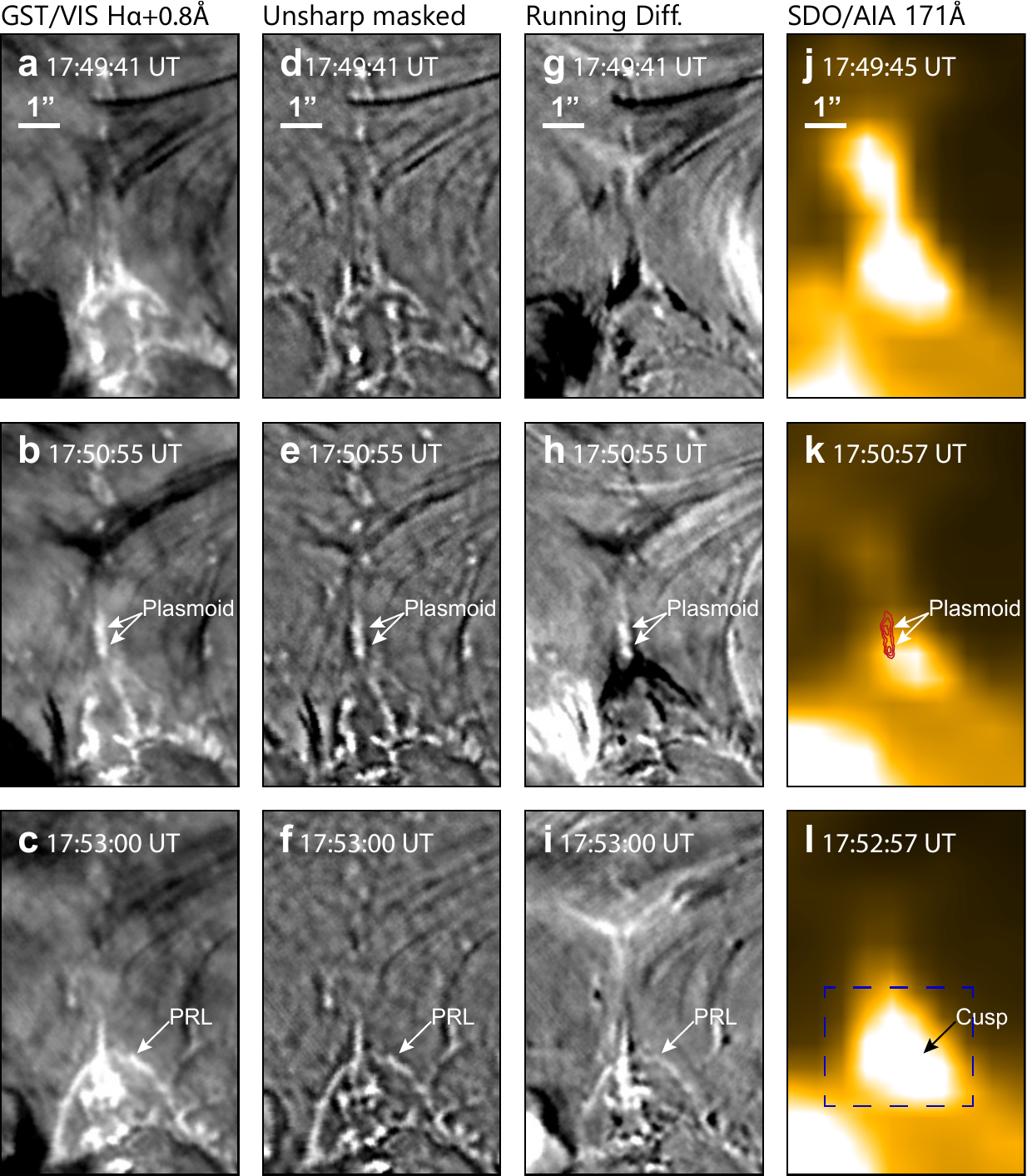}
\end{interactive}
\caption{ The evolutionary images of the current sheet. From left to right are the GST/VIS H$\alpha$+0.8 \AA\ images, the unsharp masked version of the GST/VIS H$\alpha$+0.8 \AA\ images, the running difference of the GST/VIS H$\alpha$+0.8 \AA\ images, and the SDO/AIA 171 \AA\ images. Time runs up to down. The red isolines in (k) are traced from (b). The blue rectangle in (l) denotes the area where the DEM analysis is calculated. The animation comprehensively illustrates the magnetic reconnection process of a minor solar flare, incorporating H$\alpha$ images from GST/VIS and 171 Å images from SDO/AIA, covering the period from 19:31:04 UT to 19:58:22 UT, with a duration of about 27 minutes. This static image only displays three key frames from the animation. An animated version of this figure is available.
\label{fig:general}}
\end{figure}

\begin{figure}[ht!]
\plotone{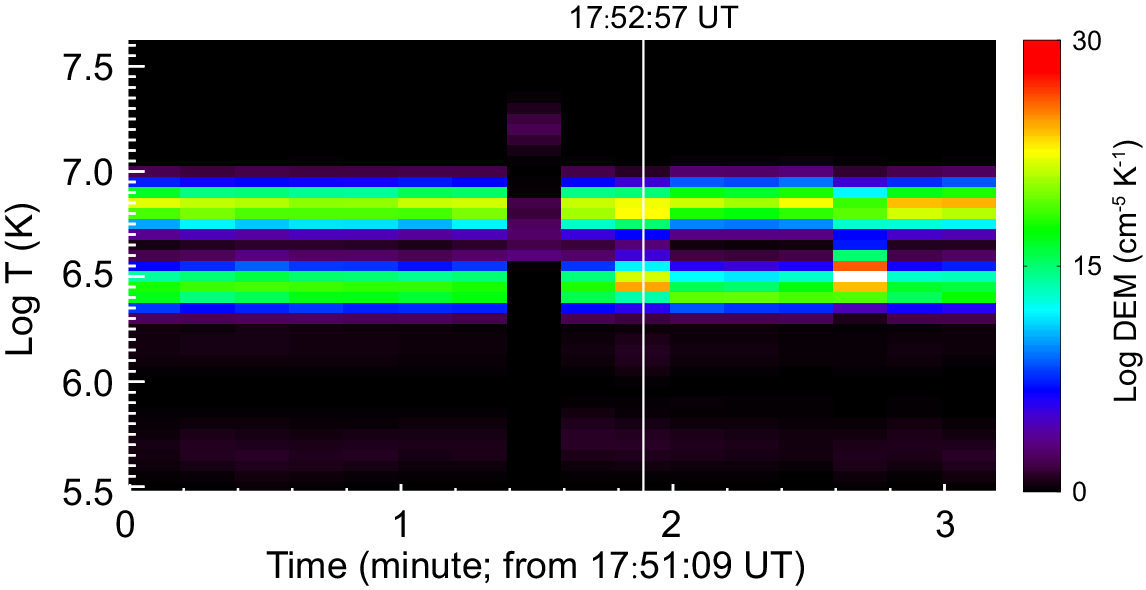}
\caption{Time-series plot of DEM analysis, where color indicates the distribution of radiative intensity, the x-axis represents the timeline of the reconnection event, and the y-axis represents lg T. The white vertical line highlighted in the plot signifies the moment when the plasmoid-like structures collide with the PRL.
\label{fig:general}}
\end{figure}

\section{Results} \label{sec:style}
\subsection{Observations}

The GST, with its 1.6-meter aperture, is one of the largest ground-based telescope worldwide, providing continuous data to the solar physics community \citep{Cao2010}. Visible Imaging Spectrometer (VIS) of GST uses an H$\alpha$ Lyot filter to capture narrow-band H$\alpha$ images (H$\alpha$ 6563 \AA \space; bandwidth: 0.25 \AA \space) with a pixel size of 0.0289" and a cadence of 25 seconds. Additionally, this study utilizes the AIA on the SDO, where AIA Level 1 data is calibrated using \texttt{"aia\_prep.pro"} in Solar Software (SSW). The SDO/AIA includes 7 extreme ultraviolet (EUV) channels (94 \AA \space to 335 \AA \space) and 2 ultraviolet (UV) channels (1600 \AA \space and 1700 \AA \space), with pixel sizes at 0.6" and cadences of 12 seconds for EUV and 24 seconds for UV data.\\

The prominent reconnection event occurred in NOAA AR 13085. We conducted a comprehensive analysis using high-resolution H$\alpha$ imaging data from the GST and extreme ultraviolet (EUV) imaging data from the AIA. We examined the evolution of the region near the current sheet and utilized NLFFF extrapolation based on solar photospheric magnetic field data from the Solar Dynamics Observatory's Helioseismic and Magnetic Imager (SDO/HMI) to assist in determining the approximate magnetic field connectivity of the active region.\\

Fig.1 shows the overview of the ROI. The observed reconnection located at the AR 13085 (Fig.1a). The GST/VIS H$\alpha$ observation clearly shows the magnetic structure of the reconnection region, which was in the X shape. Here the NLFFF extrapolation of the ROI was used to show the magnetic connectivity of the reconnection region. As shown by the extrapolated result (Figs.1c and 1d), the reconnection region consisted of four different loops connecting four different magnetic polarities. The length of the current sheet as shown in Figs.1b and 1d is about 2 Mm. The focus of this study is mainly on verifying the plasmoid-mediated chromospheric reconnection of the ROI, and investigating the secondary heating effect of the plasmoid.\\

Fig.2 presents the H$\alpha$ +0.8 \AA \space and AIA 171 \AA \space images of the reconnection region at three different times, the appearance, downward movement and dissapearance of the plasmoids are revealed, the EUV brightening of the magnetic cusp is concurrently shown. At around 17:49 UT, GST/VIS H$\alpha$+0.8 image clearly displays the reconnection region of interest, characterized by the upper absorbed cusp, the lower bright cusp (or post-reconnected loop; PRL) and the mid elongated current sheet (Fig.2a). About 1 minute later, two plasmoids with a size about 150 km appeared in the current sheet, as shown in Figs.2b, 2e and 2h. The appearance of the plasmoids suggests that reconnection was during the stage with plasmoid instability. The plasmoids then moved downward, and collided with the lower cusp (see online video). Meanwhile, the PRL was apparently lighted (Figs.2c, 2f and 2i), the strong 171 \AA \space brightening appeared at around the colliding area shown as the cups shape (Fig.2), the emissions in the lower temperature bands are also enhanced at this area. The intensified emission in the H$\alpha$+0.8 and 171 \AA \space images suggest that the local plasma was heated. The high spatio-temporal correlation between the collision of the plasmoid and the intensified emission of the H$\alpha$+0.8 \AA \space and 171 \AA \space indicates that plasmoids can heat the surrounding plasma during its collision. By tracing the plasmoid structures in current sheets through time-difference plots, it was discovered that the downward velocity of the plasmoid, at the moment it could first be identified, was 24 km$\cdot$s$^{-1}$, which then gradually decreased to 2.4 km$\cdot$s$^{-1}$. \\

We employ the method proposed by \cite{Su2018} to conduct a Differential Emission Measure (DEM) analysis in the PRL region (The blue rectangular area in Fig.2l ), utilizing observations from six AIA bands from 17:51:09 UT to 17:54:21 UT. The average results are presented in Fig.3, where the x-axis corresponds to time, and the y-axis to lg T. The distribution of colors indicates the average intensity of radiation in the analysis area, with a white vertical line in Fig.3 marking the time when the observed plasma blobs contact the PRL. Notably, following the contact of the plasma blobs with the loops, an evident enhancement in radiation around log T = 6.3 - 6.8 is observed, possibly caused by the interaction between the plasmoid and the PRL. Moreover, this event likely transcends the chromosphere, transition region, and reaches into the lower corona. The plasmoid-like structures observed in the wings of the H$\alpha$ line may also have counterparts in the lower corona, although they remain elusive due to the current resolution limitations.\\

\begin{figure}
\centering
\includegraphics[width=1.\textwidth]{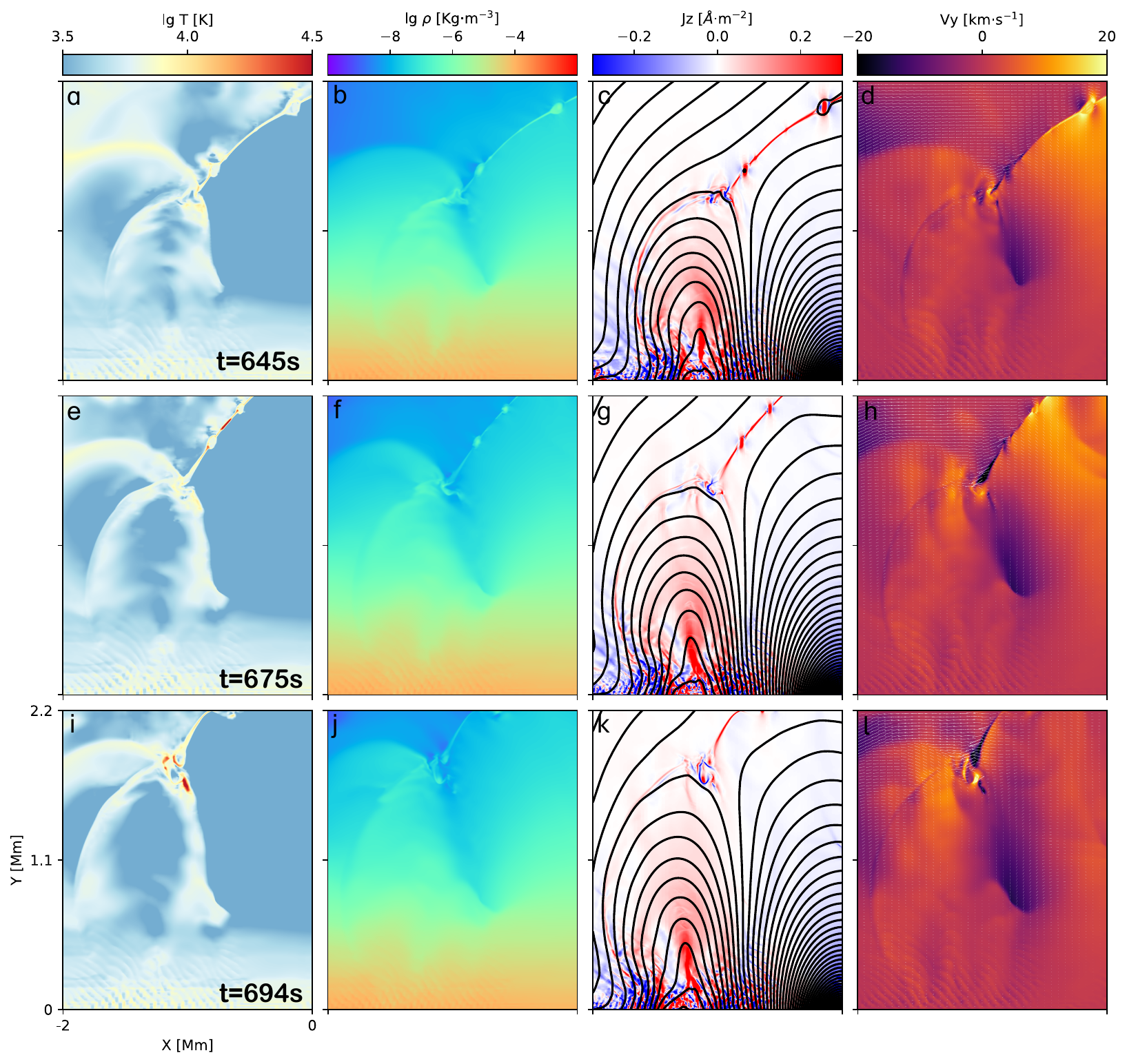}
\caption{General view of simulation results at different moments. Columns 1, 2, 3 and 4 display the temperature, density, current density ($J_{z}$) and v$_y$ distributions for three different times, respectively. The black solid lines in column 3 represent the magnetic field lines. The white arrows in column 4 represent the velocity fields.
\label{fig:general}}
\end{figure}

\subsection{MHD simulations}

We performed the high resolution 2.5D MHD simulations to reproduce a similar reconnection process and confirm the plasmoid instability can indeed happens in such an environment. A modified version of the NIRVANA code \citep{Ziegler2008,Ziegler2011} is applied. The MHD equations solved in this simulation can be referenced from our previous work\citep{Ni2021,Cheng2024}.\\

In order to simulate the more realistic low solar atmosphere, this simulation incorporates the processes of radiative cooling and partial ionization in the lower atmosphere. In the photosphere and lower chromosphere, we adopt the radiative cooling model proposed by \cite{Abbett2012}, along with the Saha equation to determine the ionization degree of the plasma environment. In the mid to upper chromosphere, we utilize the approximate radiative cooling and hydrogen ionization models provided by \cite{Carlsson2012}, In this work, we only simulate a reconnection process in the partially ionized low solar atmosphere and the corona part is excluded.\\

The simulation box is from -8 to 8 Mm in the x-direction and from 0 to 2.2 Mm in the y-direction (vertical the solar surface), which covers the domain from the solar surface to the upper chromosphere. The Adaptive Mesh Refinement (AMR) level is set to 4, hence the smallest grid size is about 2.6 km and 0.7 km in the two directions, respectively. Initial atmospheric plasma distributions are based on interpolated data from the C7 model\citep{Avrett2008}. We set the initial uniform inclined magnetic field and the time-dependent bottom boundary condition results in the emerging magnetic arch \citep{Ni2021}. Then, the emerging magnetic arch gradually reconnects with the background magnetic fields with opposite direction, an inclined elongated current sheet is formed and the plasmoid instability starts to appear during the later stage. Though the whole simulation is very similar as that in the previous work \citep{Cheng2024},  we present the results during a much later stage in this work.\\

Fig.4 presents the distributions of temperature, plasma density, current density, magnetic field lines and velocity in y direction at three different times. The distributions of magnetic field lines reveal a pronounced PRL in the lower left part of the simulation domain (Figs.4c, 4g, 4k), with a height ranging between 0.5Mm to 1.5Mm. The slightly higher density and temperature in this loop compared to the surrounding environments will cause the enhanced emissions in H$\alpha$ wing and core. It corresponds to the PRL observed below the reconnection site in Fig.1. In Fig.4, multiple plasmoids are visible within the elongated current sheet, and the interaction and merging of these plasmoids with the reconnection outflows lead to a significant temperature increase in the PRL region as shown in Fig.4i. The interaction between two plasmoids inside the current sheet also cause noticeable local temperature increases. The simulation clearly demonstrates the formation of plasmoids within the current sheet and their interactions with the PRL, agreeing well with the observations presented in the last subsection. The length scales of the current sheet and plasmoids in the simulations are also similar as those from observational results. The temperature and plasma density distributions as shown in Fig.4 are nonuniform in the current sheet and the loop regions, which can explain the enhanced emissions observed in multiple wave bands. Nevertheless, our simulations only represent part of the phenomenon that is located below the corona.\\

\section{Discussion} \label{sec:floats}

\begin{figure}[ht!]
\plotone{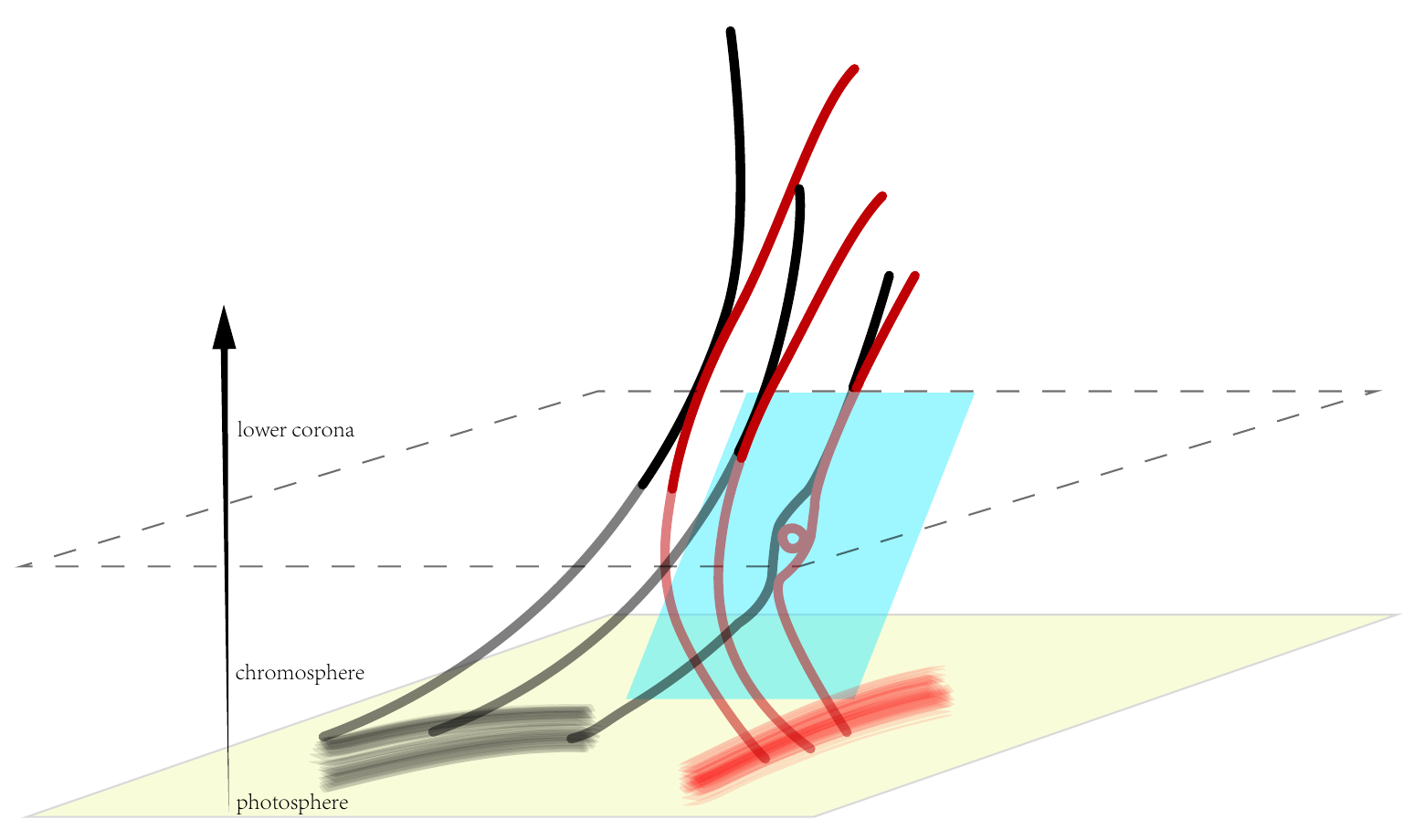}
\caption{The cartoon of the reconnection event. The bottom plane symbolizes the photospheric layer, with black and red blocks denoting regions of differing magnetic polarities. Magnetic field lines of corresponding polarities are illustrated by black and red lines, respectively. Black arrows indicate the heights of these regions. The light green plane represents the area covered by the 2.5D MHD simulation.
\label{fig:general}}
\end{figure}

Combining observational analysis, magnetic field extrapolation, DEM analysis and MHD simulation results, we can construct a scenario for this fast magnetic reconnection event, also simply illustrated by Fig.5. Initially, this small scale brightening event is clearly visible in H$\alpha$ line core and wing images, accompanied by responses in the AIA ultraviolet and extreme ultraviolet bands. The magnetic field line distributions from the NLFFF extrapolations further suggest that the brightening event is corresponding to the magnetic reconnection process. These results indicate that the reconnection occurs in the lower solar atmosphere, namely the chromosphere and transition region, with some structures extending into the lower corona. However, plasmoid structures and their interaction with the PRL are only observed in the H$\alpha$ wing images. We analyzed the movement of plasmoids, noting an initial speed of 24 km$\cdot$s$^{-1}$ that decreased to 2.4 km$\cdot$s$^{-1}$ upon contact with the PRL. When the plasmoids collide with and merge into the PRL, the area of the PRL is compressed and heated, as shown in Fig.4i. This interaction results in a noticeable enhancement in the emissions observed in the AIA 171 Å and other lower temperature bands, which is also supported by the DEM analysis presented in Fig.3. This process is similar to the observations of large-scale flaring activities\citep{Takasao2012}, they observed the interactions of the plasmoids with the PRL at a higher altitude in the corona. The length scale of the current sheet and plasmoids observed in high temperature AIA bands in their work are more than one order of magnitude larger than those shown in this work. We demonstrated that a similar phenomenon can also happen in the much deeper solar atmosphere in the partially ionized plasma. \\

Furthermore, our simulation results clearly demonstrate the formations of plasmoids in the chromospheric plasma environment and their subsequent interaction with the PRL leading to further temperature increases, consistent with the evolution of the small scale reconnection event from observations. The core results are revealed through our 2.5D MHD simulations, though the 3D full comparison with observations can not be performed. The simulation acts as a slice that presents the lower part of the entire reconnection process, indicated by the light green plane in Fig.5.\\


The observations of the current sheet structure, the downward outflows, the following enhanced emissions at around the PRL region, the NLFFF extrapolation results of magnetic filed, and our simulations with a similar process, provide substantial and credible evidences to confirm the small brightening event is caused by a small scale reconnection process. These features are very similar as those observed in a larger scale corona flare. e.g., \cite{Hou2021} presented a very similar reconnection configuration as this work, except that the reconnection current sheet in their work is much longer and mostly located in the corona. These results help us to  better comprehend solar eruptions and activities, and are consistent with the unified reconnection model for solar flares with different length scales from the lower chromosphere to the corona \citep[e.g.,][]{Shibata1996}.  Future solar telescopes with higher resolution  are expected to resolve smaller-scale nanoflares, which are crucial for understanding atmospheric heating.\\


Plasmoid instability in the reconnecting current sheets is considered as one of the main fast reconnection mechanism for explaining the explosive energy release process in the Universe. Numerous MHD and PIC simulations in fully ionized plasmas have proved that such an unstable reconnecting current sheet can exist in different length scales in the past thirty years\citep[e.g.,][]{Biskamp1993,Bhattacharjee2009,Daughton2009,Ni2010,Shepherd2010}. The pasmoids have been detected in a magnetic reconnection process in the laboratory experiments \citep[e.g.,][]{Ping2023} and magnetosphere \citep[e.g.,][]{Wang2016}, the plasmoid-like structures have also been observed in the solar corona through high resolution observations \citep[e.g.,][]{Lin2005, Takasao2012, ZhangQM2014,Hou2021}. The previous single-fluid and multi-fluids MHD simulations demonstrate that plasmoids instability can also appear in weakly ionized plasmas \citep[e.g.,][]{Leake2012,Ni2015,Ni2018}, but direct evidences from laboratory experiments or observations are very rare. The recent observations \citep{Rouppe2023} indicated that the bright blobs in H$\alpha$ wing images are possibly corresponding to the plasmoid-like structures in the deep solar atmosphere. In this work, the clear current sheet structure, the reconnection magnetic field topology, the downward plasmoid-like blobs in H$\alpha$ wing and the following brighting post flare loop comprehensively provide strong evidences for plasmoid mediated reconnection in partially ionized plasmas. \\

\section*{}
This research is supported by the Strategic Priority Research Program of the Chinese Academy of Sciences with Grant No. XDB0560000; the National Key R$\&$D Program of China No. 2022YFF0503003 (2022YFF0503000); the National Key R$\&$D Program of China No. 2022YFF0503804(2022YFF0503800); the NSFC Grants 12373060 and 11933009; the outstanding member of the Youth Innovation Promotion Association CAS (No. Y2021024 ); the Basic Research of Yunnan Province in China with Grant 202401AS070044; the Yunling Talent Project for the Youth; the Yunling Scholar Project of the Yunnan Province and the Yunnan Province Scientist Workshop of Solar Physics; Yunnan Key Laboratory of Solar Physics and Space Science under the number 202205AG070009;; The numerical calculations and data analysis have been done on Hefei advanced computing center and on the Computational Solar Physics Laboratory of Yunnan Observatories.

\bibliography{sample631}{}
\bibliographystyle{aasjournal}



\end{document}